\documentclass[aps,prl,twocolumn,superscriptaddress,showpacs,amsmath,amssymb]{revtex4} 
\usepackage{graphicx}
\usepackage{epsfig} 
\begin{document}

\newcommand{\io}{i\omega_n}
\newcommand{\cdag}{c^{\dag}}
\newcommand{\fdag}{f^{\dag}}
\newcommand{\dagga}{{\phantom{\dagger}}}
\newcommand{\m}[1]{\mathcal{#1}}
\newcommand{\ta}[2]{\tau^{#1}_{#2}}
\newcommand{\tb}[2]{\tau^{'#1}_{#2}}
\newcommand{\al}[2]{\alpha^{#1}_{#2}}
\newcommand{\ab}[2]{\alpha^{'#1}_{#2}}
\newcommand{\x}[1]{x_{#1}^{a}}
\newcommand{\y}[1]{x_{#1}^{'a}}
\newcommand{\cc}[1]{c_{a\ab{a}{#1}}(\tb{a}{#1})}
\newcommand{\cd}[1]{c^{\dag}_{a\al{a}{#1}}(\ta{a}{#1})}
\newcommand{\DD}[1]{\Delta^a_{\al{a}{#1}\ab{a}{#1}}(\ta{a}{#1},\tb{a}{#1})}
\newcommand{\DDD}[2]{\Delta^a_{\al{a}{#1}\ab{a}{#2}}(\ta{a}{#1},\tb{a}{#2})}
\newcommand{\f}[1]{\m{#1}\left(\m{C}\right)}
\newcommand{\eps}{\varepsilon}
\newcommand{\eqn}[1]{(\ref{#1})}

\title{Real-Time Diagrammatic Monte Carlo for Nonequilibrium Quantum Transport}
\author{Marco Schir\'o}
\affiliation{International School for Advanced Studies (SISSA), and CRS Democritos, CNR-INFM,
Via Beirut 2-4, I-34014 Trieste, Italy} 
\author{Michele Fabrizio} 
\affiliation{International School for Advanced Studies (SISSA), and CRS Democritos, CNR-INFM,
Via Beirut 2-4, I-34014 Trieste, Italy}
\affiliation{The Abdus Salam International Centre for Theoretical Physics 
(ICTP), P.O.Box 586, I-34014 Trieste, Italy} 
 
\date{\today} 
\begin{abstract}
We propose a novel approach to nonequilibrium real-time dynamics of quantum impurities models coupled to biased non-interacting leads, 
such as those relevant to quantum transport in nanoscale molecular devices. The method is based on a Diagrammatic Monte Carlo sampling 
of the \emph{real-time} perturbation theory along the Keldysh contour. We benchmark the method on a non-interacting resonant level 
model and, as a first non-trivial application, we study zero temperature non-equilibrium transport through a vibrating molecule.
\end{abstract}
\maketitle

\textit{Introduction.}  Recent advances in nanotechnology have made it possible to contact microscopic quantum objects, like  
artificial atoms (quantum dots), molecules or quantum wires, with metallic electrodes, opening the route towards very promising  
nanoelectronical devices~\cite{Tao_nanonature,Cuniberti_book}. Following the first discovery of the Kondo effect in 
quantum dots~\cite{GoldhaberGordon_prl98}, a lot of challenging experiments have been carried out contacting, e.g., 
single-molecules~\cite{Park_single_molecule,parks-C60} 
or coupled quantum dots~\cite{coupled_QD} to metallic leads.
In the meantime, these impressive experimental developments have 
raised a lot of novel and interesting physical questions, so that electronic transport through nano-systems has become one of the 
frontier fields in condensed matter physics. 
Indeed, the possibility contact microscopic objects is particulary 
intriguing as it allows to study quantum transport in a regime where the tunneling rate becomes comparable 
or even smaller than other energy scales, like the electron-electron repulsion or the energy of 
atomic displacements, which may lead to interesting non-equilibrium effects.~\cite{Ralph,H2O}

These experimental progresses urgently ask for developing efficient non-perturbative theoretical tools 
to treat out-of-equilibrium phenomena. The simplest way to model nonequilibrium transport in nanodevices is through 
a quantum impurity model, namely a set of discrete levels $a$ 
(with creation operators $c^{\dag}_{a\sigma}$, $\sigma$ being the spin), 
mimicking a quantum dot or a molecule, coupled to two baths of non interacting electrons 
(with creation operators $f^{\dag}_{k\,\sigma\alpha}$), labeled by some quantum number $k$, which account for the metallic leads. 
The leads ($\alpha=L,R$) are kept at different chemical potentials $\mu_{L}-\mu_R=eV$. As a consequence the general Hamiltonian 
may read 
\begin{eqnarray}\label{eqn:H_qim} 
\m{H}&=& \sum_{\alpha=L,R}\sum_{k\,\sigma}\,\epsilon_{k\,\alpha}\,f^{\dag}_{k\,\sigma\alpha}\,f^\dagga_{k\,\sigma\alpha} 
+\m{H}_{loc}\left[\cdag_{a\sigma},c_{a\sigma}\right]\nonumber\\
&&+\sum_{k\,a\,\alpha\,\sigma}\,\bigg(V_{k\,a\,\alpha}\,f^{\dag}_{k\,\sigma\alpha}c^\dagga_{a\sigma} +h.c.\bigg)\,.
\end{eqnarray}
The local Hamiltonian $\m{H}_{loc}$ accounts for all the physics on the quantum impurity, including possibly 
vibrational degrees of freedoms, and, because of the discrete set of levels, could in principle be diagonalized exactly. 
However, the hybridization to the reservoirs makes the problem untreatable unless in simple cases.  
Furthermore, the finite bias, which drives the system out of equilibrium, rules out the possibility 
to apply all the powerful tools developed during last decades, like 
Numerical Renormalization Group (NRG)~\cite{Pruschke_RMP} and the recent   
Diagrammatic Monte Carlo method (DiagMC)~\cite{ctqmc_Rubtsov,ctqmc_Werner}.\\
Standard approaches to nonequilibrium are usually based on Keldysh perturbation theory~\cite{Konig_prb}, which is 
analytically feasible only to lowest orders. In order to access to the most interesting 
intermediate-coupling regime several numerical methods has been proposed~\cite{tnrg_Anders,iter_Egger,Rabani}.\\
In this Letter we present a novel real-time DiagMC approach to nonequilibrium transport in quantum impurity models. 
The method is based on a stochastic sampling of the Keldysh diagrams generated by the perturbative expansion in the 
tunnelling. It does not require any discretization of the time evolution, hence it provides a very accurate 
description of the dynamics. 
We benchmark the method on a non-interacting resonant level and then, as a first non-trivial application, 
we compute the inelastic tunneling spectrum in the resonant level coupled to a local vibrational mode.\\
\textit{Formulation.}  To set up the method, we consider an initially decoupled system, made by the isolated impurity 
and the two leads, each assumed to be at equilibrium with its own reservoir at chemical potential $\mu_{\alpha}$. 
At time $t>0$, we switch-on the hybridization in (\ref{eqn:H_qim}) 
and let the system evolve with the full hamiltonian $\m{H}$. Given an initial density matrix, $\rho_0$, we want to compute 
average values of physical operators evolved in real-time from $0$ to $t$. 
The final goal is to succeed following the dynamics till the steady state, so to compute 
experimentally relevant quantities like the differential conductance $dI/dV$.\\
\begin{figure}
\begin{center}
\includegraphics[width=5.0cm]{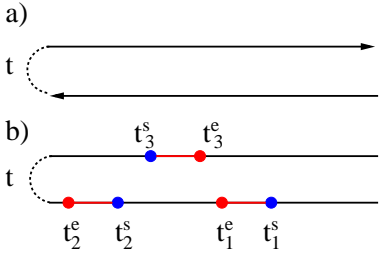}
\caption{\label{fig:contour} a) Keldysh contour $\m{K}$ used in real-time diagMC. Under time-ordering $T_{\m{K}}$ 
creation/annihilation operators are displaced along the contour time-ordered as shown by the arrows. 
b) An example of a configuration 
with $k=3$ segments, each starting at $t^{s}_i$ and ending at $t^{e}_i$, $i=1,2,3$. Here blue/red dots stands for 
annihilation/creation operators of the initially occupied level, while red segments indicate how the vertices are connected 
by the hybridization functions $\Delta_{\m{K}}(t^e,t^s)$ in the particular configuration shown.}
\end{center}
\end{figure}
Real-time quantum dynamics can be generally represented as evolution along the so-called Keldysh contour $\m{K}$, plotted in 
Fig.~\ref{fig:contour}, made of two branches winding around the real-time axis from $0$ to $t$ (lower branch) and back from 
$t$ to $0$ (upper branch). In this perspective the average value of any operator $\m{O}$ can be written as 
\begin{equation}\label{eqn:O_K}
\langle\m{O}\left(t\right)\rangle=Tr\left\{\rho_0\:T_\m{K}\left(e^{-i\int_{\m{K}}\,d\tau\,\m{H}\left(\tau\right)}\m{O}\right)\right\},
\end{equation}
where the trace is over the lead and impurity degrees of freedom and $T_{\m{K}}$ denotes time-ordering   
along the Keldysh contour. The main idea of the approach is to expand the evolution operator~(\ref{eqn:O_K}) in powers of the 
hybridization and trace-out exactly the leads degrees of freedom.
The expansion obtained in that way looks as the natural generalization of the diagrammatic expansion of Ref.~\cite{ctqmc_Werner} 
to the $\m{K}$ contour, which is required to deal with nonequilibrium effects. The resulting diagrams are then sampled with a 
Monte Carlo algorithm that we shall discuss in detail. To show how the method works, we start by considering a spinless biased 
resonant level model (RLM), namely a single fermionic energy level $\eps_d$ driven out of equilibrium by an applied bias 
$eV=\mu_L-\mu_R$ between the two leads. Given an initial density matrix $\rho_0= \rho_{leads}\otimes \rho_{imp}$, with 
$\rho_{leads}$ describing the two uncoupled leads each at equilibrium with its own chemical potential and    
$\rho_{imp}=c^\dagger c$ -- the impurity is initially occupied --  we are interested in computing, for example, 
the real-time evolution of the RLM population $n=c^{\dag}c$. To this extent, we expand the Keldysh evolution operator in powers 
of the hybrization between the bath and the impurity and trace out the lead degrees of freedom, which are non-interacting. 
The resulting expansion reads
\begin{widetext}
\begin{eqnarray}
\langle n(t) \rangle&=&\sum_{k=0}^{\infty}\sum_{n=0}^{k}(-1)^k\, 
\int_0^{t_{k}^s} dt_k^e \int_0^{t_{k-1}^e} dt_k^s \dots 
\int_0^{t_{k-n}^s} dt_{k-n}^e \int_0^t\,dt_{k-n}^s\nonumber\\
&&\int_0^t dt_{k-n-1}^e \int_0^{t_{k-n-1}^e} dt_{k-n-1}^s \dots \int_0^{t_2^s}dt_1^e \int_0^{t_1^e}dt_{1}^s\,
\m{D}_k\left(t_1^e,\dots,t_k^e|t_1^s,\dots,t_k^s\right)\,\m{L}_k\left(t_1^e,\dots,t_k^e|t_1^s,\dots,t_k^s\right)
\end{eqnarray}
\end{widetext}
Here $\m{D}_k$ is the outcome of tracing the lead degrees of freedom and can be expressed in closed form as the 
determinant of a $k\times k$ matrix $\m{M}^{-1}$,
\begin{equation}\label{eqn:D_k}
\m{D}\left(t_1^e,\dots,t_k^e|t_1^s,\dots,t_k^s\right) = det\,\left(\m{M}^{-1}\right), 
\end{equation}
whose entries are the Keldysh hybridization functions
\begin{equation}\label{eqn:delta}
\m{M}^{-1}_{ij}=i\Delta_{\m{K}}\left(t_i^e,t_j^s\right)\,s\left(t_i^e,t_j^s\right), 
\end{equation}
which we define as 
\begin{equation}
\Delta_{\m{K}}\left(t^e,t^s\right)=\sum_{k\alpha}\vert V_{k\alpha}\vert^2
\langle T_{\m{K}}\left(f^\dagga_{k\alpha}\left(t^e\right)f^{\dag}_{k\alpha}\left(t^s\right)\right)\rangle\,.
\end{equation}
Here we adopt the standard definition of the Keldysh Green's functions~\cite{Rammer_book_noneq}, namely
we consider $t^s,t^e$ as living on the contour $\m{K}$. We note the additional sign $s\left(t^e,t^s\right)$, 
which is negative when the two times are on opposite branches and positive otherwise. While the determinant $\m{D}_k$ 
properly accounts for the effects associated to the leads, the function $\m{L}_k$ involves only the impurity 
degrees of freedom and can be generally written as a trace over the initial impurity density matrix, namely
\begin{equation}\label{eqn:T_k}
\m{L}_k=Tr\bigg\{\rho_{imp}\,T_{\m{K}}\left(c^{\dag}(t^e_k)c(t_k^s)\dots 
c^{\dag}(t^e_1)c(t_1^s) n(t)\right)\bigg\}.
\end{equation}
The expansion thus obtained admits a natural representation in terms of a collection of $k$ segments, 
$t\in\left[t^s_i,t^e_i\right]$, or equivalently $k-1$ anti-segments, $t\in\left[t^e_i,t^s_{i+1}\right]$,   
properly ordered along the contour $\m{K}$ and connected in all possible ways by the Keldysh hybridization functions 
$\Delta_{\m{K}}(t^e,t^s)$. An example of such a configuration for $k=3$ is plotted in fig.~\ref{fig:contour}. 
It is worth noticing that a similar expansion can be carried out also for other quantities 
like e.g. the average current $\langle I_{\alpha}(t)\rangle$ or the noise $S(t)=\langle I_{\alpha}(t)I_{\alpha}(0)\rangle$,
resulting into a very general approach.\\
\textit{Algorithm. } As usual in DiagMC~\cite{ctqmc_Rubtsov,ctqmc_Werner} we view the perturbative expansion as a sum over 
configurations $\m{C}$, i.e. diagrams with $k$ segments placed along the contour $\m{K}$. 
In order to get an efficient sampling scheme three basic updates are implemented: adding/removing a segment, 
adding/removing an antisegment or shifting a segment end-point. 
We accept/reject a new configuration 
according to detailed balance prescription. In the actual simulation, we store and update the matrix $\m{M}$ 
defined in (\ref{eqn:delta}), which is the only quantity required to compute Metropolis 
acceptance ratios~\cite{ctqmc_Rubtsov}.\\
\textit{Benchmark. } We benchmark the method in the biased spinless RLM, which can be exactly solved hence being 
the simplest test for a real-time diagMC calculation. In fact, the main issue 
when sampling real-time quantum dynamics is the so called sign-problem, namely the exponential increase of 
relative errors in the Monte Carlo estimate of any observable in the infinite-size (or zero temperature) limit. 
This is well known in thermal-equilibrium (imaginary-time) Monte Carlo simulations of fermionic systems. 
In this perspective, diagrammatic Monte Carlo has been proved to be a generic sign-tolerant approach which can deal   
with sign-alterning series and even take advantage from them~\cite{Prokofev_sign}. 
How far this approach can be pushed, in particular in studying real-time dynamics, is the main scope of this work.
To this extent, we compute the occupation number $\langle n(t)\rangle$ as a function of time $t$ for different values 
of the level position $\eps_d$ both in equilibrium, $eV=0$, as well as out of equilibrium, $eV\neq0$, and compare the exact 
results with diagMC data. In this simple case, we expect that a single energy scale, namely the level 
broadening $\Gamma=\pi\sum_k\,|V_k|^2\delta(\eps_k)$, controls the approach to the steady-state. As can be seen 
from Fig.~\ref{fig1}, this is indeed confirmed by diagMC calculation which perfectly matches the exact solution. 
Indeed, we are able to resolve both the short-time decay from the initial configuration as well as the approach to the steady state. 
We note that a finite applied bias, $eV\neq0$, cuts off the Keldysh evolution operator~(\ref{eqn:O_K}) 
(the steady state is reached earlier than at $eV=0$), as pointed out by Ref.~\cite{Mitra_Millis_prl05}, 
thus making the expansion more convergent. As mentioned previously, within the present 
approach one can easily measure the current flowing 
thorugh the impurity $I(t)=\frac{1}{2}\langle I_L(t)-I_R(t)\rangle$ on a fine real-time grid in a very efficient way, 
which is also shown in Fig.~\ref{fig1}. It is worth mentioning that within our Keldysh diagMC we are able to reach a 
true \textit{non-equilibrium} steady state with a finite value of the current, due to the infinite size of the bath.
Dissipation occurs entirely within the fermionic reservoir and we do not need to include any ficticious bosonic bath 
to reach a steady state.\\
\begin{figure}
\begin{center}
\includegraphics[width=8.9cm]{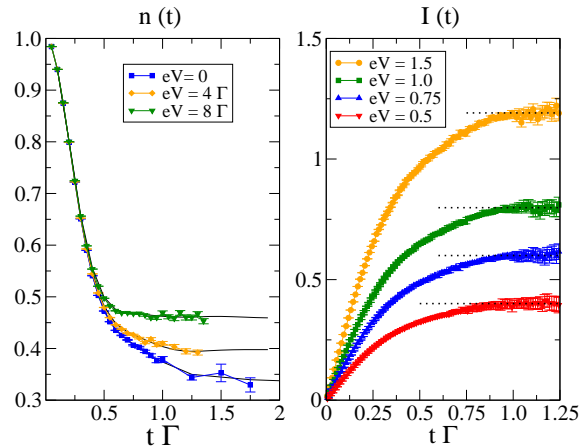}
\caption{\label{fig1} Zero-temperature real-time dynamics for different bias values $eV$ of the dot population, 
$\langle n(t)\rangle$, from an initially occupied dot (left panel) and of the  current, $\langle I(t)\rangle$ (right panel)  
of the resonant level model. 
DiagMC results (dots) are compared with the exact solution (full line). We take $\epsilon_d=\Gamma$ and 
consider a flat density of states in the leads with an half-bandwidth $10\Gamma$.}
\end{center}
\end{figure}
\textit{Nonequilibrium Transport through a single-molecule. }As a first non trivial application we consider a simple model of a 
molecular conductor, namely a spinless fermionic level coupled to Holstein phonon. The local Hamiltonian reads
\begin{equation}\label{eqn:H_loc_holst}
\m{H}_{loc}(n) = \frac{\omega_0}{2} (x^2+p^2) + gx(n-\frac{1}{2}) + \varepsilon_d(n-\frac{1}{2})
\end{equation}
where $\omega_0$ is the phonon frequency (with displacement $x$ and its conjugate variable $p$), 
$\eps_d$ is the energy of the level and $n$ its occupancy and finally $g$ the electron-phonon coupling. 
Our Keldysh diagMC can be naturally extendend to 
include local phonons, the only difference appears in the trace over local degrees of freedom (\ref{eqn:T_k}), 
which now involves fermionic operators evolved according to the hamiltonian (\ref{eqn:H_loc_holst}) for the electron-phonon subsystem. 
This trace can be evaluated analytically by observing that the local Hamiltonians with different level occupancy $n=0,1$ 
are related one to the other by a unitary transformation, $\m{H}_{loc}(0)+\epsilon_d=U^{\dag}\m{H}_{loc}(1)U$,   
with $U=\exp\left(igp/\omega_0\right)$. It follows that the bosonic contribution to the local trace reduces to 
the following bosonic correlation-function 
\begin{equation}\label{eqn:T_K_ph}
\m{L}^{ph}_k=Tr\Bigg(\rho_{ph}\,U^\dagger(t^e_k)\,U(t^s_k)\,U^\dagger(t^e_{k-1})\dots U(t^s_1)\Bigg),
\end{equation}
which can be easily evaluated analytically for most common initial phonon density matrices  
$\rho_{ph}$, which we assume the equlibrium distribution at zero temperature. 
In \eqn{eqn:T_K_ph} $U(t)$ and $U^\dagger(t)$ are the unitary operators evolved 
with $\m{H}_{loc}(1)$, and we have assumed the level initially occupied. 
Therefore local vibrational degrees of freedom can be included in our sampling scheme at any order in the coupling constant $g$, 
allowing to extract by Keldysh DiagMC non-perturbative results in the electron-phonon coupling.
\begin{figure}
\begin{center}
\includegraphics[width=8.9cm]{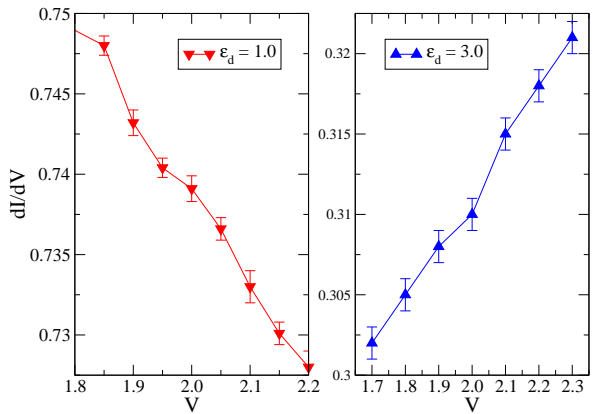}
\caption{\label{fig2} Zero-temperature differential conductance $dI/dV$ (in unit of $e^2/h$) for bias values 
$eV$ around $\omega_0=2.0\Gamma$. Electron-phonon coupling is $g=0.5\omega_0$. We consider two different values of the 
fermionic level position $\eps_d$ in order to reproduce the crossover from reducted (step-down) to enhanced (step-up) conductance.}
\end{center}
\end{figure}\\
The coupling to molecular vibrations is known to significantly affect inelastic electron tunneling.~\cite{Galperin} When the bias 
hits a vibrational frequency, the differential conductance $dI/dV$ changes sharply. Experimentally it is observed 
that $dI/dV$ increases in the tunneling regime, but decreases in the opposite case of a high transmission barrier. 
Although simple physical arguments can be invoked~\cite{Jaklevic,Persson,Agrait} to explain this phenomenon, theoretical 
calculations within the Keldysh formalism have so bar been restricted to the lowest orders in the electron-phonon 
coupling.~\cite{vega,paulsson:2008,Egger} In the simple resonant level model that we are considering, the  
perturbative calculations predict that $dI/dV$, at bias $eV=\omega_0$, should decrease if the 
zero-bias conductance $G> 0.5$, in units of the unitary value (in our spinless case $e^2/h$), 
otherwise should increase.~\cite{vega,paulsson:2008,Egger} Evidences in favor of this results have been 
very recently found by Tal and coworkers with H$_2$O molecules bridging a Pt break-junction. Since the Keldysh diagMC is,  
as mentioned, non perturbative in the electron-phonon coupling, it offers the possibility to verify the above 
theoretical predictions.        
We model the two regimes of $G\gtrless 0.5$ by two different values of the level position 
$\eps_d =1$ and 3, the former closer to resonance than the latter, and compute directly the differential conductance $dI/dV$.   
As can be seen from Fig.~\ref{fig2} 
a step-down or step-up features do appear around the threshold for vibronic excitations, $eV\simeq\omega_0$,  
when the zero bias conductance is greater or lower than $0.5$, respectively, 
in agreement with perturbative results.~\cite{vega,paulsson:2008,Egger} We also note that the step is not as abrupt 
as found in perturbation theory~\cite{vega,paulsson:2008,Egger},  
likely signaling a significant phonon damping.  

In summary, we have introduced a novel non perturbative approach to nonequilibrium quantum transport in nanoscopic conductors. 
The method is based on a Diagrammatic Monte Carlo sampling of the \textit{real-time} perturbation theory 
in the impurity-leads hybridization, performed along the Keldysh contour required to treat non equilibrium effects. 
In spite of the oscillating nature of the real-time quantum evolution, we are able to follow the dynamics starting from an 
arbitrary initial preparation up to the steady state. This is primarily due to the combined effects of infinite leads and of the 
applied bias, which cut-off the Keldysh evolution operator, but also to the capability of the algorithm to cope with sign-problem. 
As a first application we have studied zero temperature non linear transport through a simple model of a molecular conductor. 
Being a completely general method, it can be in principle used to study any discrete quantum system bridging two non-interacting 
conducting leads, providing a new tool to study quantum transport in nanoscopic devices.

\begin{thebibliography}{25}
\expandafter\ifx\csname natexlab\endcsname\relax\def\natexlab#1{#1}\fi
\expandafter\ifx\csname bibnamefont\endcsname\relax
  \def\bibnamefont#1{#1}\fi
\expandafter\ifx\csname bibfnamefont\endcsname\relax
  \def\bibfnamefont#1{#1}\fi
\expandafter\ifx\csname citenamefont\endcsname\relax
  \def\citenamefont#1{#1}\fi
\expandafter\ifx\csname url\endcsname\relax
  \def\url#1{\texttt{#1}}\fi
\expandafter\ifx\csname urlprefix\endcsname\relax\def\urlprefix{URL }\fi
\providecommand{\bibinfo}[2]{#2}
\providecommand{\eprint}[2][]{\url{#2}}

\bibitem[{\citenamefont{Tao}(2006)}]{Tao_nanonature}
\bibinfo{author}{\bibfnamefont{N.~J.} \bibnamefont{Tao}},
  \bibinfo{journal}{Nature Nanotechnology} \textbf{\bibinfo{volume}{1}},
  \bibinfo{eid}{173-181} (\bibinfo{year}{2006}).

\bibitem[{\citenamefont{Cuniberti et~al.}(2005)\citenamefont{Cuniberti, Fagas,
  and Richter}}]{Cuniberti_book}
\bibinfo{author}{\bibfnamefont{G.}~\bibnamefont{Cuniberti}},
  \bibinfo{author}{\bibfnamefont{G.}~\bibnamefont{Fagas}}, \bibnamefont{and}
  \bibinfo{author}{\bibfnamefont{K.}~\bibnamefont{Richter}},
  \emph{\bibinfo{title}{Introducing Molecular Electronic}}
  (\bibinfo{publisher}{Springer Berlin and Heidelberg}, \bibinfo{year}{2005}).

\bibitem[{\citenamefont{Goldhaber-Gordon
  et~al.}(1998)\citenamefont{Goldhaber-Gordon, G\"ores, Kastner, Shtrikman,
  Mahalu, and Meirav}}]{GoldhaberGordon_prl98}
\bibinfo{author}{\bibfnamefont{D.}~\bibnamefont{Goldhaber-Gordon}},
  \bibinfo{author}{\bibfnamefont{J.}~\bibnamefont{G\"ores}},
  \bibinfo{author}{\bibfnamefont{M.~A.} \bibnamefont{Kastner}},
  \bibinfo{author}{\bibfnamefont{H.}~\bibnamefont{Shtrikman}},
  \bibinfo{author}{\bibfnamefont{D.}~\bibnamefont{Mahalu}}, \bibnamefont{and}
  \bibinfo{author}{\bibfnamefont{U.}~\bibnamefont{Meirav}},
  \bibinfo{journal}{Phys. Rev. Lett.} \textbf{\bibinfo{volume}{81}},
  \bibinfo{pages}{5225} (\bibinfo{year}{1998}).

\bibitem[{\citenamefont{Park et~al.}(2000)\citenamefont{Park, Park, Lim,
  Anderson, ALivisatos, and Mcuen}}]{Park_single_molecule}
\bibinfo{author}{\bibfnamefont{H.}~\bibnamefont{Park}},
  \bibinfo{author}{\bibfnamefont{J.}~\bibnamefont{Park}},
  \bibinfo{author}{\bibfnamefont{A.}~\bibnamefont{Lim}},
  \bibinfo{author}{\bibfnamefont{E.}~\bibnamefont{Anderson}},
  \bibinfo{author}{\bibfnamefont{A.}~\bibnamefont{ALivisatos}},
  \bibnamefont{and} \bibinfo{author}{\bibfnamefont{P.}~\bibnamefont{Mcuen}},
  \bibinfo{journal}{Nature} \textbf{\bibinfo{volume}{407}}, \bibinfo{eid}{57}
  (\bibinfo{year}{2000}).

\bibitem[{\citenamefont{Parks et~al.}(2007)\citenamefont{Parks, Champagne,
  Hutchison, Flores-Torres, Abruna, and Ralph}}]{parks-C60}
\bibinfo{author}{\bibfnamefont{J.~J.} \bibnamefont{Parks}},
  \bibinfo{author}{\bibfnamefont{A.~R.} \bibnamefont{Champagne}},
  \bibinfo{author}{\bibfnamefont{G.~R.} \bibnamefont{Hutchison}},
  \bibinfo{author}{\bibfnamefont{S.}~\bibnamefont{Flores-Torres}},
  \bibinfo{author}{\bibfnamefont{H.~D.} \bibnamefont{Abruna}},
  \bibnamefont{and} \bibinfo{author}{\bibfnamefont{D.~C.} \bibnamefont{Ralph}},
  \bibinfo{journal}{Phys. Rev. Lett.} \textbf{\bibinfo{volume}{99}},
  \bibinfo{eid}{026601} (\bibinfo{year}{2007}).

\bibitem[{\citenamefont{Stinaff et~al.}(2006)\citenamefont{Stinaff, Scheibner,
  Bracker, Ponomarev, Korenev, Ware, Doty, Reinecke, and Gammon}}]{coupled_QD}
\bibinfo{author}{\bibfnamefont{E.~A.} \bibnamefont{Stinaff}},
  \bibinfo{author}{\bibfnamefont{M.}~\bibnamefont{Scheibner}},
  \bibinfo{author}{\bibfnamefont{A.~S.} \bibnamefont{Bracker}},
  \bibinfo{author}{\bibfnamefont{I.~V.} \bibnamefont{Ponomarev}},
  \bibinfo{author}{\bibfnamefont{V.~L.} \bibnamefont{Korenev}},
  \bibinfo{author}{\bibfnamefont{M.~E.} \bibnamefont{Ware}},
  \bibinfo{author}{\bibfnamefont{M.~F.} \bibnamefont{Doty}},
  \bibinfo{author}{\bibfnamefont{T.~L.} \bibnamefont{Reinecke}},
  \bibnamefont{and} \bibinfo{author}{\bibfnamefont{D.}~\bibnamefont{Gammon}},
  \bibinfo{journal}{Science} \textbf{\bibinfo{volume}{311}},
  \bibinfo{pages}{636} (\bibinfo{year}{2006}).

\bibitem[{\citenamefont{Shi et~al.}(2007)\citenamefont{Shi, Bolotin, Kuemmeth,
  and Ralph}}]{Ralph}
\bibinfo{author}{\bibfnamefont{S.-F.} \bibnamefont{Shi}},
  \bibinfo{author}{\bibfnamefont{K.~I.} \bibnamefont{Bolotin}},
  \bibinfo{author}{\bibfnamefont{F.}~\bibnamefont{Kuemmeth}}, \bibnamefont{and}
  \bibinfo{author}{\bibfnamefont{D.~C.} \bibnamefont{Ralph}},
  \bibinfo{journal}{Phys. Rev. B} \textbf{\bibinfo{volume}{76}},
  \bibinfo{eid}{184438} (\bibinfo{year}{2007}).

\bibitem[{\citenamefont{Tal et~al.}(2008)\citenamefont{Tal, Krieger, Leerink,
  and van Ruitenbeek}}]{H2O}
\bibinfo{author}{\bibfnamefont{O.}~\bibnamefont{Tal}},
  \bibinfo{author}{\bibfnamefont{M.}~\bibnamefont{Krieger}},
  \bibinfo{author}{\bibfnamefont{B.}~\bibnamefont{Leerink}}, \bibnamefont{and}
  \bibinfo{author}{\bibfnamefont{J.~M.} \bibnamefont{van Ruitenbeek}},
  \bibinfo{journal}{Phys. Rev. Lett.} \textbf{\bibinfo{volume}{100}},
  \bibinfo{eid}{196804} (\bibinfo{year}{2008}).

\bibitem[{\citenamefont{Bulla et~al.}(2008)\citenamefont{Bulla, Costi, and
  Pruschke}}]{Pruschke_RMP}
\bibinfo{author}{\bibfnamefont{R.}~\bibnamefont{Bulla}},
  \bibinfo{author}{\bibfnamefont{T.~A.} \bibnamefont{Costi}}, \bibnamefont{and}
  \bibinfo{author}{\bibfnamefont{T.}~\bibnamefont{Pruschke}},
  \bibinfo{journal}{Rev. Mod. Phys.} \textbf{\bibinfo{volume}{80}},
  \bibinfo{eid}{395} (\bibinfo{year}{2008}).

\bibitem[{\citenamefont{Rubtsov et~al.}(2005)\citenamefont{Rubtsov, Savkin, and
  Lichtenstein}}]{ctqmc_Rubtsov}
\bibinfo{author}{\bibfnamefont{A.~N.} \bibnamefont{Rubtsov}},
  \bibinfo{author}{\bibfnamefont{V.~V.} \bibnamefont{Savkin}},
  \bibnamefont{and} \bibinfo{author}{\bibfnamefont{A.~I.}
  \bibnamefont{Lichtenstein}}, \bibinfo{journal}{Phys. Rev. B}
  \textbf{\bibinfo{volume}{72}}, \bibinfo{eid}{035122} (\bibinfo{year}{2005}).

\bibitem[{\citenamefont{Werner et~al.}(2006)\citenamefont{Werner, Comanac, de'
  Medici, Troyer, and Millis}}]{ctqmc_Werner}
\bibinfo{author}{\bibfnamefont{P.}~\bibnamefont{Werner}},
  \bibinfo{author}{\bibfnamefont{A.}~\bibnamefont{Comanac}},
  \bibinfo{author}{\bibfnamefont{L.}~\bibnamefont{de' Medici}},
  \bibinfo{author}{\bibfnamefont{M.}~\bibnamefont{Troyer}}, \bibnamefont{and}
  \bibinfo{author}{\bibfnamefont{A.~J.} \bibnamefont{Millis}},
  \bibinfo{journal}{Phys. Rev. Lett.} \textbf{\bibinfo{volume}{97}},
  \bibinfo{eid}{076405} (\bibinfo{year}{2006}).

\bibitem[{\citenamefont{Governale et~al.}(2008)\citenamefont{Governale, Pala,
  and K\"{o}nig}}]{Konig_prb}
\bibinfo{author}{\bibfnamefont{M.}~\bibnamefont{Governale}},
  \bibinfo{author}{\bibfnamefont{M.~G.} \bibnamefont{Pala}}, \bibnamefont{and}
  \bibinfo{author}{\bibfnamefont{J.}~\bibnamefont{K\"{o}nig}},
  \bibinfo{journal}{Phys. Rev. B} \textbf{\bibinfo{volume}{77}},
  \bibinfo{eid}{134513} (\bibinfo{year}{2008}).

\bibitem[{\citenamefont{Anders and Schiller}(2005)}]{tnrg_Anders}
\bibinfo{author}{\bibfnamefont{F.~B.} \bibnamefont{Anders}} \bibnamefont{and}
  \bibinfo{author}{\bibfnamefont{A.}~\bibnamefont{Schiller}},
  \bibinfo{journal}{Phys. Rev. Lett.} \textbf{\bibinfo{volume}{95}},
  \bibinfo{eid}{196801} (\bibinfo{year}{2005}).

\bibitem[{\citenamefont{Weiss et~al.}(2008)\citenamefont{Weiss, Eckel,
  Thorwart, and Egger}}]{iter_Egger}
\bibinfo{author}{\bibfnamefont{S.}~\bibnamefont{Weiss}},
  \bibinfo{author}{\bibfnamefont{J.}~\bibnamefont{Eckel}},
  \bibinfo{author}{\bibfnamefont{M.}~\bibnamefont{Thorwart}}, \bibnamefont{and}
  \bibinfo{author}{\bibfnamefont{R.}~\bibnamefont{Egger}},
  \bibinfo{journal}{Phys. Rev. B} \textbf{\bibinfo{volume}{77}},
  \bibinfo{eid}{195316} (\bibinfo{year}{2008}).

\bibitem[{\citenamefont{M\"{u}hlbacher and Rabani}(2008)}]{Rabani}
\bibinfo{author}{\bibfnamefont{L.}~\bibnamefont{M\"{u}hlbacher}}
  \bibnamefont{and} \bibinfo{author}{\bibfnamefont{E.}~\bibnamefont{Rabani}},
  \bibinfo{journal}{Phys. Rev. Lett.} \textbf{\bibinfo{volume}{100}},
  \bibinfo{eid}{176403} (\bibinfo{year}{2008}).

\bibitem[{\citenamefont{Rammer}(2008)}]{Rammer_book_noneq}
\bibinfo{author}{\bibfnamefont{J.}~\bibnamefont{Rammer}},
  \emph{\bibinfo{title}{Quantum Field Theory of Nonequilibrium States}}
  (\bibinfo{publisher}{Cambridge University Press}, \bibinfo{year}{2008}).

\bibitem[{\citenamefont{Prokof'ev and Svistunov}(2007)}]{Prokofev_sign}
\bibinfo{author}{\bibfnamefont{N.}~\bibnamefont{Prokof'ev}} \bibnamefont{and}
  \bibinfo{author}{\bibfnamefont{B.}~\bibnamefont{Svistunov}},
  \bibinfo{journal}{Phys. Rev. Lett.} \textbf{\bibinfo{volume}{99}},
  \bibinfo{eid}{250201} (\bibinfo{year}{2007}).

\bibitem[{\citenamefont{Mitra et~al.}(2005)\citenamefont{Mitra, Aleiner, and
  Millis}}]{Mitra_Millis_prl05}
\bibinfo{author}{\bibfnamefont{A.}~\bibnamefont{Mitra}},
  \bibinfo{author}{\bibfnamefont{I.}~\bibnamefont{Aleiner}}, \bibnamefont{and}
  \bibinfo{author}{\bibfnamefont{A.~J.} \bibnamefont{Millis}},
  \bibinfo{journal}{Phys. Rev. Lett.} \textbf{\bibinfo{volume}{94}},
  \bibinfo{eid}{076404} (\bibinfo{year}{2005}).

\bibitem[{\citenamefont{Galperin et~al.}(2007)\citenamefont{Galperin, Ratner,
  and Nitzan}}]{Galperin}
\bibinfo{author}{\bibfnamefont{M.}~\bibnamefont{Galperin}},
  \bibinfo{author}{\bibfnamefont{M.}~\bibnamefont{Ratner}}, \bibnamefont{and}
  \bibinfo{author}{\bibfnamefont{A.}~\bibnamefont{Nitzan}},
  \bibinfo{journal}{J.Phys.: Condens. Matter} \textbf{\bibinfo{volume}{19}},
  \bibinfo{eid}{103201} (\bibinfo{year}{2007}).

\bibitem[{\citenamefont{Jaklevic and Lambe}(1966)}]{Jaklevic}
\bibinfo{author}{\bibfnamefont{R.~C.} \bibnamefont{Jaklevic}} \bibnamefont{and}
  \bibinfo{author}{\bibfnamefont{J.}~\bibnamefont{Lambe}},
  \bibinfo{journal}{Phys. Rev. Lett.} \textbf{\bibinfo{volume}{17}},
  \bibinfo{pages}{1139} (\bibinfo{year}{1966}).

\bibitem[{\citenamefont{Persson}(1988)}]{Persson}
\bibinfo{author}{\bibfnamefont{B.}~\bibnamefont{Persson}},
  \bibinfo{journal}{Phys. Scr.} \textbf{\bibinfo{volume}{38}},
  \bibinfo{pages}{282} (\bibinfo{year}{1988}).

\bibitem[{\citenamefont{Agra\"{\i}t et~al.}(2002)\citenamefont{Agra\"{\i}t,
  Untiedt, Rubio-Bollinger, and Viera}}]{Agrait}
\bibinfo{author}{\bibfnamefont{N.}~\bibnamefont{Agra\"{\i}t}},
  \bibinfo{author}{\bibfnamefont{C.}~\bibnamefont{Untiedt}},
  \bibinfo{author}{\bibfnamefont{G.}~\bibnamefont{Rubio-Bollinger}},
  \bibnamefont{and} \bibinfo{author}{\bibfnamefont{S.}~\bibnamefont{Viera}},
  \bibinfo{journal}{Chem. Phys.} \textbf{\bibinfo{volume}{281}},
  \bibinfo{pages}{231} (\bibinfo{year}{2002}).

\bibitem[{\citenamefont{de~la Vega et~al.}(2006)\citenamefont{de~la Vega,
  Martin-Rodero, Agrait, and Yeyati}}]{vega}
\bibinfo{author}{\bibfnamefont{L.}~\bibnamefont{de~la Vega}},
  \bibinfo{author}{\bibfnamefont{A.}~\bibnamefont{Martin-Rodero}},
  \bibinfo{author}{\bibfnamefont{N.}~\bibnamefont{Agrait}}, \bibnamefont{and}
  \bibinfo{author}{\bibfnamefont{A.~L.} \bibnamefont{Yeyati}},
  \bibinfo{journal}{Phys. Rev.B} \textbf{\bibinfo{volume}{73}},
  \bibinfo{eid}{075428} (\bibinfo{year}{2006}).

\bibitem[{\citenamefont{Paulsson et~al.}(2008)\citenamefont{Paulsson,
  Frederiksen, Ueba, Lorente, and Brandbyge}}]{paulsson:2008}
\bibinfo{author}{\bibfnamefont{M.}~\bibnamefont{Paulsson}},
  \bibinfo{author}{\bibfnamefont{T.}~\bibnamefont{Frederiksen}},
  \bibinfo{author}{\bibfnamefont{H.}~\bibnamefont{Ueba}},
  \bibinfo{author}{\bibfnamefont{N.}~\bibnamefont{Lorente}}, \bibnamefont{and}
  \bibinfo{author}{\bibfnamefont{M.}~\bibnamefont{Brandbyge}},
  \bibinfo{journal}{Phys. Rev. Lett.} \textbf{\bibinfo{volume}{100}},
  \bibinfo{eid}{226604} (\bibinfo{year}{2008}).

\bibitem[{\citenamefont{Egger and Gogolin}(2008)}]{Egger}
\bibinfo{author}{\bibfnamefont{R.}~\bibnamefont{Egger}} \bibnamefont{and}
  \bibinfo{author}{\bibfnamefont{A.}~\bibnamefont{Gogolin}},
  \bibinfo{journal}{Phys. Rev. B} \textbf{\bibinfo{volume}{77}},
  \bibinfo{eid}{113405} (\bibinfo{year}{2008}).

\end{thebibliography}

\end{document}